\documentclass[aps,pra,superscriptaddress,twocolumn,showpacs]{revtex4}
\usepackage{graphics}
\usepackage{epsfig}
\usepackage{graphicx}
\usepackage{amsthm} % for theorems
\usepackage{amsbsy} % for bold symbols (\bf doesn't work for Greek letters)
\usepackage{dcolumn}   % Align table columns on decimal point
\usepackage{verbatim}   % useful for program listings
\usepackage{color}     	 % use if color is used in text
\usepackage{subfigure}  % use for side-by-side figures
\usepackage{amsmath,amsfonts,amssymb,graphics,graphicx,epsfig,color,times,natbib}

\theoremstyle{remark}

\theoremstyle{definition}

\newcommand{\be}{\begin{equation}}
\newcommand{\ee}{\end{equation}}
\newcommand{\ba}{\begin{eqnarray}}
\newcommand{\ea}{\end{eqnarray}}
\newcommand{\ban}{\begin{eqnarray*}}
\newcommand{\ean}{\end{eqnarray*}}

\newcommand{\bra}[1]{\left\langle#1\right|} % } These resise
\newcommand{\ket}[1]{\left|#1\right\rangle} % }
\newcommand{\bristolmath}{\affiliation{Department of Mathematics, University of Bristol, Bristol BS8 1TW,
United Kingdom}}
\newcommand{\bristol}{\affiliation{H.H. Wills Physics Laboratory, University of Bristol, Bristol BS8 1TL, United Kingdom}}
\newcommand{\gdansk}{\affiliation{Institute of Theoretical Physics and Astrophysics, University of
Gdansk, 80-952 Gdansk, Poland}}
\begin{document}
\title{Semi-device-independent security of one-way quantum key distribution}
\author{Marcin Paw{\l}owski}\bristolmath\gdansk
\author{Nicolas Brunner}\bristol
%\pacs{75.10.Pq,	03.65.Ud, 03.67.-a}
\date{\today}

\begin{abstract}
By testing nonlocality, the security of entanglement-based quantum key distribution (QKD) can be enhanced to being 'device-independent'. Here we ask whether such a strong form of security could also be established for one-way (prepare and measure) QKD. While fully device-independent security is impossible, we show that security can be guaranteed against individual attacks in a semi-device-independent scenario. In the latter, the devices used by the trusted parties are non-characterized, but the dimensionality of the quantum systems used in the protocol is assumed to be bounded. Our security proof relies on the analogies between one-way QKD, dimension witnesses and random-access codes.
\end{abstract}
\maketitle

The aim of quantum cryptography \cite{review} is to warrant security against an eavesdropper solely limited by the laws of quantum mechanics. However, any quantum key distribution (QKD) scheme relies on an additional assumption which concerns information leakage out of the laboratories of Alice and Bob. Specifically, both parties must be free to choose which measurement they perform in each run of the protocol, and this choice of measurement, as well as the outcome of this measurement, should remain unknown to the eavesdropper. Indeed if the eavesdropper has access to the lab of Alice or Bob, then security cannot be guaranteed.

Apart from these basic requirements, standard security proofs of QKD \cite{shorpreskill} also assume that Alice and Bob have an excellent control on the quantum states and measurements used in the protocol. This assumption is however hard to justify in practice, where devices always feature some level of imperfection. Moreover this assumption turns out to be crucial, as nicely illustrated in Ref. \cite{acin06}. There it was shown that the security of the Bennett-Brassard (BB84) protocol \cite {bb84} is entirely compromised if Alice and Bob use 4-dimensional states instead of qubits---as usual security proofs always assume. It is however possible to avoid this requirement by basing the security on nonlocality. Specifically, by checking for the violation of a Bell inequality \cite{ekert}, Alice and Bob can ensure that they share nonlocal correlations, in which case security can be guaranteed without having any detailed knowledge on the functioning of the cryptographical devices \cite{mayers,bhk}. This is 'device-independent' (DI) QKD \cite{DI} (see also \cite{DI2}).

%This renewed recently interest in entanglement based QKD protocols in which the security is assessed by checking for the violation of a Bell inequality. By testing NL, independent of QM and description of the devices used in the protocol, one can establish device-independent security.

The promise of a higher level of security, as well as the recently demonstrated attacks on actual QKD systems \cite{lydersen}, have motivated research towards the practical implementation of DI-QKD. Despite recent progress \cite{gisin10}, this remains a challenging problem. Moreover, the fact that DI-QKD is based on nonlocality strongly suggests that only entanglement based protocols are suitable for obtaining this stronger notion of security. However, almost none of the practical QKD systems, in particular none of the commercially available ones, use entanglement; they all operate in a one-way configuration, in which Alice prepares a quantum state, sends it to Bob who then performs a measurement on it (hence often called 'prepare and measure').

Here we will argue that a form of DI security---thus stronger than usual security proofs---can nevertheless be obtained for QKD protocols which do not make use of entanglement. Specifically, we shall see that in a \emph{semi-device-independent} scenario, in which the devices are non-characterized but only assumed to produce quantum systems of a given dimension, security of one-way QKD against individual attacks can be demonstrated. In particular our proof will make use of the analogy between one-way QKD protocols, dimension witnesses \cite{gallego} and random-access codes \cite{nayak}. To the best of our knowledge, our work represents the first QKD security proof that can be applied directly to the one-way configuration.

We shall start by presenting the semi-DI scenario we consider, stating clearly all assumptions we make. Then, we will consider the BB84 protocol and show that it becomes completely insecure in this context.
This will also make clear that dimension witnesses are suitable tools for tackling this problem. Next we will discuss the intimate relation existing between dimension witnesses and random-access codes \cite{wehner}. Finally we will describe a specific QKD protocol and derive, via its associated dimension witness (or random-access code), a security proof.

\begin{figure}[b]
 \begin{center}
  \includegraphics[width=5.5cm,angle=0]{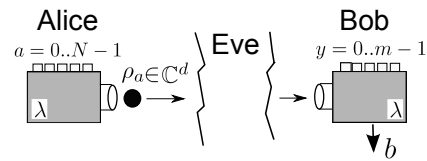}
  \caption{Semi-device-independent one-way QKD.}
 \label{setup}
 \end{center}
\end{figure}

\section{Preliminaries} 

In a one-way QKD scheme Alice encodes classical information in a quantum system, which she sends to Bob via a quantum channel. Bob then performs a measurement on the system, from which he decodes some information. After repeating these operations many times, Alice and Bob estimate the error rate (by revealing randomly chosen bits from the raw key) which leads to an upper bound on Eve's information. Finally Alice and Bob perform classical post-processing---error-correction, privacy amplification---to extract the sifted key on which Eve has arbitrarily small information.

Here we shall work in a semi-DI scenario. That is, we will assume that the (relevant) Hilbert space dimension of the quantum systems is known \footnote{Here we mean the dimension of the relevant part of the Hilbert space, that is, the degrees of freedom that are correlated with the classical information encoded by Alice.}, but that the quantum preparations and measurements are non-characterized. It will thus be convenient to describe the devices of Alice and Bob by black boxes. Specifically Alice's black box is a 'state preparator'. Alice has the freedom to choose among a certain set of preparations $\rho_{a} \in \mathbb{C}^d$ with $a \in \{0,..,N-1\}$, but knows nothing about these quantum states apart from their dimensionality $d$. We also assume that Alice's preparations $\rho_{a}$ are unentangled from Eve---note that if Alice's preparations were entangled with Eve's system, then the communication capacity would be effectively doubled using dense coding. Bob's device is a measurement black-box. He can choose to perform a measurement $M_y$ with $y\in \{0,...,m-1\}$ and gets the outcome $b\in \{0,...,k-1\}$. The measurement operators $M_y$ are non-characterized; note that Eve could in principle send a system of arbitrary dimension to Bob. The boxes may also feature shared classical variables $\lambda$, known to Eve, but uncorrelated from the choice of preparation made by Alice and the choice of measurement made by Bob.

After repeating this procedure many times, Alice and Bob can estimate the probability distribution (or data table \cite{harrigan})
\ba P(b|a,y)=\text{tr}(\rho_a M_y^b),\ea
which denotes the probability of Bob finding outcome $b$ when he performed measurement $M_y$, and Alice prepared $\rho_a$. Our goal will be to show that, in some cases, the security of a given protocol against a quantum eavesdropper can be guaranteed solely from its associated data table $P(b|a,y)$. The security is thus semi-DI, in the sense that we do not require any knowledge on how the data table $P(b|a,y)$ was obtained, except from the fact that the device of Alice emits quantum systems of a given dimension.

Here we will restrict ourselves to individual attacks, in which Eve attacks independently each system sent by Alice (using the same strategy) and measures her system before the classical post-processing \cite{review}. Indeed we also need to make the basic assumption about information leakage from the devices. That is, no information about the inputs and output (i.e. $a$, $y$ and $b$) leaks out of the boxes to Eve.

\section{Dimension witnesses} 

At this point one can already see a first requirement for obtaining semi-DI security for a given protocol. Suppose Alice's device prepares $d$-dimensional quantum systems. Then it must be impossible to reproduce the quantum data table with classical systems of dimension $d$. If not, then it could have been the case that Alice's device emits orthogonal quantum states (or equivalently classical states) from which Eve can get full information. Thus, full DI security, that is where no assumption is made on the Hilbert space dimension, is impossible, since every data table can be reproduced using classical systems of sufficient dimension.

It turns out that a simple method for establishing lower bounds on the dimension of classical systems necessary to reproduce a given data table was recently developed in Ref. \cite{gallego}. More precisely, they authors devised 'dimension witnesses', of the form
\ba\label{DW} \sum_{a,y,b} w_{aby}P(b|a,y) \leq C_d, \ea
which can be thought of as Bell-type inequalities for data tables. Here the bound $C_d$ denotes the maximal value of the left-hand-side polynomial obtainable when Alice's device prepares classical $d$-dimensional systems. Interestingly, $d$-dimensional classical dimension witnesses can be violated by $d$-dimensional quantum systems, thus indicating that certain quantum data tables cannot be reproduced using classical states of the same dimension. Below we will make use of this 'quantum advantage'. We will consider a simple dimension witness which provides a separation between qubits and bits. In particular we will show how this witness can be naturally understood as a random-access code, which will allow us to prove semi-DI security of the corresponding QKD protocol.

From now we will now focus on the case where Alice's device prepares 2-dimensional quantum systems, and restrict to four preparations ($N=4$) indexed by two bits $a_0,a_1$. Bob's device can perform two binary measurements ($m=2$, $k=2$). For the rest of the paper it will be convenient to use expectation values of the form
\ba E_{a_0a_1,y} = P(b=0|a_0a_1,y). \ea
Thus every experiment corresponds to a data table, given by a vector $\vec{E}=\{E_{a_0a_1,y}\}_{a_0a_1,y}$ of $Nm=8$ correlators.

First, we would like to characterize the set of data tables (i.e. the set of vectors $\vec{E}$) which can be obtained when Alice's box emits classical bits. We follow the geometrical methods of Ref. \cite{gallego}. The set of interest to us is a polytope (in an 8-dimensional space). Its facets are (tight) 2-dimensional classical witnesses; that is inequalities of the form \eqref{DW} with $d=2$ (note that here probabilities are simply replaced by correlators).
It turns out that there are only two types of witnesses in the case. The first is a straightforward extension \footnote{Note that the witness $I_3$ of \cite{gallego} uses only $N=3$ preparations. Here we obtain the lifting of $I_3$ to the case $N=4$; the fourth preparation does simply not appear into the expression.} of the witness $I_3$ of Ref. \cite{gallego}. The second is of the following form:
\ba\label{I_QKD}\nonumber S &=&  +E_{00,0} + E_{00,1} + E_{01,0} - E_{01,1} \\ & &   -E_{10,0} + E_{10,1}  -E_{11,0} - E_{11,1}  \leq 2\ea
This witness will be our main tool to assess the security of one-way QKD protocols.

\emph{BB84 is not secure.} As a warm-up, it is instructive to consider first the case of the BB84 protocol. In this case, the four preparations of Alice are given by \ba\label{preps}\nonumber & &\rho_{00} = \ket{0}\bra{0} \quad , \quad \rho_{11} = \ket{1}\bra{1} \\ & &\rho_{10} = \ket{+}\bra{+} \quad ,\quad  \rho_{01} = \ket{-}\bra{-}. \ea
Here $\ket{0}$ and $\ket{1}$ are the eigenstates of the Pauli matrix $\sigma_z$, and $\ket{\pm}=(\ket{0}\pm \ket{1})/\sqrt{2}$ are the eigenstates of $\sigma_x$. The two measurements of Bob are given by $M_0=\sigma_z$ and $M_1=\sigma_x$.

Thus the corresponding data table is given by $E_{00,0}=\text{tr}(\rho_{00}\sigma_z)=1$, and similarly $E_{10,1}=1$, $E_{01,1}=E_{11,0}=0$, and $E_{00,1}=E_{01,0}=E_{10,0}=E_{11,1}=1/2$. Thus the BB84 data table achieves $S=2$ and thus does not violate the witness \eqref{I_QKD}---note that it also satisfies the witness $I_3$ of \cite{gallego} as well as all symmetries---which indicates that it can be reproduced by sending one classical bit when the boxes of Alice and Bob share randomness. Note that this a peculiarity of the BB84 data table \footnote{It turns out that any deviation from the BB84 data table, obtained from modifying one (or more) preparations and/or one measurement, will lead to a value of $S>2$. Thus BB84 appears to be worst case in this context.}. Indeed this result also applies to any protocol using the same states and measurements as BB84, for instance the SARG protocol \cite{sarg}.

A possible strategy is the following. Alice and Bob share one random bit $\lambda$. Considering Alice's preparations, note that the bit $a_0\oplus a_1$ denotes the basis, while the bit $a_1$ denotes the encoded bit. When $\lambda=0$, Alice sends to Bob the (one bit) message $m=a_0 $. Bob, upon getting his input $y$ and the message from Alice $m$, outputs $b=m\oplus y=a_0\oplus  y$. Thus $b=a_1$ whenever Alice and Bob choose the same basis ($a_0\oplus a_1=y$), and $b\neq a_1$ when they choose a different basis ($a_0\oplus a_1 \neq y$). When $\lambda=1$, Alice sends the message $m=a_1$, and Bob outputs $b=m=a_1$. Thus we have that $a_1=b$ for any pair of basis $a_0\oplus a_1,y$. Since the shared variable $\lambda$ is unbiased, Alice and Bob reproduce the BB84 data table.

\section{Connection to random-access codes} 

To devise a secure QKD protocol in the semi-DI setup, we need to consider data tables which violate (at least) one of the dimension witness $I_3$ or $S$. Here we shall focus on the latter, which it will be useful to think of in terms of a random-access code.

Specifically, let us imagine that Alice receives two (uniformly distributed) bits $a_0$ and $a_1$. She is then allowed to send a physical system to Bob, which encodes information about her input bit string. Bob is asked to guess the $y$-th bit of Alice ($y$ is uniformly distributed as well), and thus performs a measurement on the system he received from Alice to extract this information. This is a 2-to-1 random-access code. When Alice sends one bit of classical communication, the optimal average probability for Bob to succeed is 3/4 \cite{nayak}.

The witness $S$ \eqref{I_QKD} represents a 2-to-1 random-access code. For each of her four possible input bit strings $\{a_0,a_1\}$, Alice associates a preparation $\rho_{a_0a_1}$. Upon being asked to guess bit $y$, Bob performs measurement $M_y$. The outcome of the measurement $b$ is then his guess for $a_y$.

From inspection of $S$, we see that $w_{a_0a_1,y}=(-1)^{a_y}$ (where $w_{a_0a_1,y}$ is the coefficient of the term $E_{a_0a_1,y}$), which implies that
\ba S  = \sum_{a_0,a_1,y} P(b=a_y|a_0a_1,y) -4  \ea
Thus, for a given data table, Bob's success probability
\ba P_B = \frac{1}{8} \sum_{a_0,a_1,y} P(b=a_y|a_0a_1,y) = \frac{S+4}{8} \ea
is determined by the value of the dimension witness $S$, and inversely. Indeed the inequality $P_{B}\leq 3/4$ corresponds to $S\leq 2$.
Note that the relation between dimension witnesses and random-access codes can be generalized (see also \cite{wehner} for a related approach).

It turns out that Alice and Bob can perform better at this task when using qubits. The optimal set of preparations are, for instance, obtained by having preparations \eqref{preps}, but changing Bob's measurements to $M_{0}=(\sigma_z+\sigma_x)/\sqrt{2}$ and $M_{1}=(\sigma_z-\sigma_x)/\sqrt{2}$. This choice of preparations and measurements leads to $S = 8 \cos^2{(\pi/8)}-4$ or equivalently
\ba\label{p_Q} P_B=\cos^2{(\pi/8)} \approx 0.8536. \ea
Note that this set of preparations and measurements is intimately related to the Clauser-Horne-Shimony-Holt Bell inequality (see also \cite{wehner}).

\section{Security of one-way QKD} 

The protocol is based on the preparations and measurements achieving the optimal violation of $S$ for qubits. Alice generates two random bits $a_0,a_1$ and sends the corresponding preparations $\rho_{a_0a_1}$ to Bob. Bob generates a random bit $y$ and performs measurement $M_y$ and guesses bit $a_y$. After repeating these operations a large number of times (we consider here only the asymptotic limit), Alice and Bob can estimate the data table by revealing part of their data on a public channel. By computing the value of $S$ they obtain $P_{B}$. Below we show that if $P_{B}>\frac{5+\sqrt{3}}{8}\approx 0.8415$---a value slightly lower than the optimal value using qubits \eqref{p_Q}---security is obtained.

\emph{Proof.} Csiszar and K\"orner \cite{CK} showed that Alice and Bob can obtain a secret key if $I(A:B)>I(A:E)$, where the mutual information is given by
\ba I(A:X) = \sum_j 1-h(P_X(a_{y_j})). \ea
Here $y_j$ denotes the choice of basis (or equivalently which bit of Alice party $X$ chose to guess) in the $j$-th run of the protocol, and $h(p)$ is the Shannon binary entropy. From this, one can get a sufficient condition for security given by
\ba P_B > P_E \ea
where $P_X= \frac{1}{2}(P_X(a_0)+P_X(a_1))$ denotes the average probability of guessing correctly for party $X$.

Our main ingredient will be a result derived by K\"onig \cite{konig}. Consider the set $F_n$ of all (boolean) balanced functions on $n$-bit strings---that is which return 0 for exactly half of the $2^n$ strings. Alice gets as input the $n$-bit string and Bob is asked to guess the value of a randomly (and uniformly) chosen function in $F_n$ after receiving from Alice $s$ qubits. Then the average probability for Bob to succeed is upper bounded as follows
\ba P_n\leq \frac{1}{2}\left(1+\sqrt{\frac{2^s-1}{2^n-1}} \right). \ea
For the case of interest to us, i.e. $n=2$ bits, the set of all balanced functions is $a_0$, $a_1$, $a_0 \oplus a_1$, and their negations. Clearly the optimal probability of guessing a function or its negation are equal. Thus, when Alice sends a single qubit to Bob ($s=1$), we have that
\ba\label{konig}
P_B(a_0)+P_B(a_1)+P_B(a_0\oplus a_1) \leq \frac{3}{2}\left(1+\frac{1}{\sqrt{3}} \right). \ea

Clearly the previous inequality holds also when Bob and Eve collaborate---the index $B$ is then simply replaced by $BE$---and we will make use of it in this case. Using the relations $P_{BE}(a_i)\geq P_{B}(a_i)$ and $P_{BE}(a_i)\geq P_{E}(a_i)$ and
\ba\nonumber P_{BE}(a_0\oplus a_1) &\geq& P_{BE}(a_0,a_1) \\
&\geq& P_{BE}(a_0)+P_{BE}(a_1)-1,\ea
where the second inequality follows from the sum rule, we get
\ba\nonumber & & P_{BE}(a_0)+P_{BE}(a_1)+P_{BE}(a_0\oplus a_1)  \\ & & \quad\quad \geq
2P_{B}(a_0)+2P_{E}(a_1)-1. \ea
Using \eqref{konig} we get that
\ba\label{eve}  P_{B}(a_0)+P_{E}(a_1)  \leq \frac{5+\sqrt{3}}{4}  \ea
and an analogous inequality with $a_0$ and $a_1$ interchanged. This shows that when Eve tries to guess a different bit than Bob (i.e. she measures in the wrong basis) she will necessarily disturb the statistics of Bob. From inequality \eqref{eve} and its symmetry with respect to $a_0$ and $a_1$, we get that
\ba \label{mon} P_B+P_{E} \leq \frac{5+\sqrt{3}}{4}. \ea
This implies that $P_B>P_E$ as long as
\ba\label{security} P_B>\frac{5+\sqrt{3}}{8}\approx 0.8415 \ea
as announced. For the optimal qubit preparations and measurements achieving \eqref{p_Q}, the key rate is found to be 
\ba r = I(A:B)-I(A:E) \approx 0.0581. \ea

\section{Discussion} 

We have discussed the security of one-way QKD in a semi-device-independent context. By making links to dimension witnesses and random-access codes, we showed that security against individual attacks is possible.

It is natural to ask whether this concept is relevant from a practical viewpoint. Since semi-DI QKD represents a relaxation of the assumption of standard QKD proofs, it offers several advantages, notably that no assumptions on the devices are required (apart from the fact that Alice's device emits preparations of bounded dimension), and that it can be applied directly to the one-way configuration. At this stage, our result should however be understood as a proof-of-principle. A next step would be to study robustness to imperfections (such as losses or detection efficiency) as well as against more general attacks. It would also be interesting to improve on our bound for security (which is likely to be suboptimal), and to see whether all data tables violating a classical dimension witness could offer security. In this context it might also be relevant to consider entropic quantities \cite{wehner,marco}.

A comparison to full DI QKD is also worth. Arguably the main drawback of our approach is the assumption of bounded dimensionality, as it forces us to assume that Alice's device features no side-channels from which Eve could extract information. This requirement could however be partly lifted by finding protocols where qubits offer security under the assumption that the preparations are arbitrary quantum states of higher dimensions---note that this would require protocols using more preparations.

Finally, from a more foundational point of view, it would be interesting to study the connection between semi-DI one-way QKD and DI entanglement-based QKD, in the light of the strong link that exists between nonlocality and random-access codes \cite{IC}.

\emph{Acknowledgments.} M.P. thanks Rob Spekkens and Marek {\.Z}ukowski for useful discussions. This work was supported by the UK EPSRC, EU QESSENCE and the Leverhulme Trust.


\begin{thebibliography}{19}

\bibitem{review} V. Scarani, H. Bechmann-Pasquinucci, N.J. Cerf, M. Dusek, N. L\"utkenhaus, M. Peev, Rev. Mod. Phys.
{\bf 81}, 1301 (2009).

\bibitem{shorpreskill} P.W. Shor and J. Preskill. Phys. Rev. Lett. {\bf 85}, 441 (2000); D. Mayers, JACM {\bf 48}, 351 (2001). 

\bibitem{acin06} A. Acin, N. Gisin, and L. Masanes, Phys. Rev. Lett. {\bf 97},  120405  (2006).

\bibitem{bb84} C.H. Bennett and G. Brassard, Proc. IEEE Int. Conf. Computers, Systems and Signal Processing, New York,
175 (1984).

\bibitem{ekert} A.~K. Ekert, Phys. Rev. Lett. 67, 661 (1991).

\bibitem{mayers} D. Mayers, A. Yao, Quant. Inf. Comput 4, 273 (2004).

\bibitem{bhk} J. Barrett, L. Hardy, and A. Kent, Phys. Rev. Lett. {\bf 95},  010503  (2005).

\bibitem{DI} A. Acin, N. Brunner, N. Gisin, S. Massar, S. Pironio, and V. Scarani, Phys. Rev. Lett. {\bf 98},  230501  (2007); S.~Pironio, A.~Ac\'in, N.~Brunner, N.~Gisin, S.~Massar, and V.~Scarani, New J. Phys. {\bf 11}, 045021 (2009).

\bibitem{DI2} L. Masanes, S. Pironio, and A. Acin, Nat. Comms {\bf 2}, 238 (2011); E. H\"anggi and R. Renner, arXiv:1009.1833; M. McKague, PhD Thesis, arXiv:1006.2352.

\bibitem{lydersen} L. Lydersen, C. Wiechers, C. Wittmann, D. Elser, J. Skaar, and V. Makarov, Nature Photonics {\bf 4}, 686 (2010).

\bibitem{gisin10} N. Gisin, S. Pironio, and N. Sangouard, Phys. Rev. Lett. {\bf 105}, 070501 (2010).

\bibitem{gallego} R. Gallego, N. Brunner, C. Hadley, and A. Acin, Phys. Rev. Lett. {\bf 105}, 230501 (2010).

\bibitem{nayak} A. Nayak, in Proceedings of 40th IEEE FOCS (1999), pp. 369376.

\bibitem{wehner} S. Wehner, M. Christandl, and A. C. Doherty, Phys. Rev. A {\bf 78}, 062112 (2008).

\bibitem{harrigan} N. Harrigan, T. Rudolph, and S. Aaronson, arxiv:0709.1149.

\bibitem{sarg} V. Scarani, A. Acin, G. Ribordy, and N. Gisin, Phys. Rev. Lett. {\bf 92}, 057901 (2004).

\bibitem{konig} R. K\"onig, PhD Thesis, (see Corrolary 5.2.3).

\bibitem{CK} I. Csiszar and J. K\"orner, IEEE Trans. Inf. Theory {\bf 24}, 339 (1978).

\bibitem{marco} M. Tomamichel and R. Renner, Phys. Rev. Lett. {\bf 106}, 110506 (2011).

\bibitem{IC} M. Pawlowski, T. Paterek, D. Kaszlikowski, V. Scarani, A. Winter, and M. Zukowski, Nature {\bf 461},  1101  (2009); J. Oppenheim and S. Wehner, Science {\bf 330}, 1072 (2010).



\end{thebibliography}
\end{document}